\documentclass[default,iicol]{sn-jnl}% Default with double column layout

\jyear{2021}%

\theoremstyle{thmstyleone}%
\theoremstyle{thmstyletwo}%

\theoremstyle{thmstylethree}%

\usepackage{mathtools}
\usepackage{amsmath}
\usepackage{amssymb}
\usepackage{wasysym}
\usepackage{amsfonts}
\usepackage{multirow}
\catcode`\|=12\relax
\usepackage{tikz}
\usetikzlibrary{arrows,quotes,angles}
\usepackage{physics}
\usepackage{caption}
\usepackage{yfonts}
\usepackage{euscript}
\usepackage{tablefootnote}
\usepackage{yfonts}
\usepackage{graphicx}
\usepackage{hyperref}
\usepackage{cleveref}
\usepackage{float}
\usepackage{graphicx}% Include figure files
\usepackage{dcolumn}% Align table columns on decimal point
\usepackage{bm}% bold math
\usetikzlibrary{snakes}
\usetikzlibrary{calc,decorations.markings}
\usepackage[compat=1.1.0]{tikz-feynman}
\usepackage{subfig}
\usepackage{feynmf}
\Crefname{equation}{Eq.}{Eqs.}
\Crefname{figure}{Fig.}{Figs.}
\Crefname{definition}{Def.}{Defs.}

\newcommand{\sbra}[1]{\langle #1 |}

\begin{document}

\title[Article Title]{Fundamental length scale and the bending of light in a gravitational field}

%%=============================================================%%
%% Prefix	-> \pfx{Dr}
%% GivenName	-> \fnm{Joergen W.}
%% Particle	-> \spfx{van der} -> surname prefix
%% FamilyName	-> \sur{Ploeg}
%% Suffix	-> \sfx{IV}
%% NatureName	-> \tanm{Poet Laureate} -> Title after name
%% Degrees	-> \dgr{MSc, PhD}
%% \author*[1,2]{\pfx{Dr} \fnm{Joergen W.} \spfx{van der} \sur{Ploeg} \sfx{IV} \tanm{Poet Laureate} 
%%                 \dgr{MSc, PhD}}\email{iauthor@gmail.com}
%%=============================================================%%

\author*[1,2]{\fnm{Philip} \sur{Tee}}\email{ptee2@asu.edu}

\author[3,4]{\fnm{Nosratollah} \sur{Jafari}}\email{nosrat.jafari@fai.kz}
\equalcont{These authors contributed equally to this work.}

\affil*[1]{\orgdiv{Beyond Center}, \orgname{Arizona State University}, \orgaddress{\city{Tempe}, \postcode{85287}, \state{Arizona}, \country{United States}}}

\affil[2]{\orgdiv{Department of Informatics}, \orgname{University of Sussex}, \orgaddress{\city{Falmer},  \state{Sussex}, \postcode{ BN1 9RH},\country{United Kingdom}}}

\affil[3]{\orgdiv{Department of Physics}, \orgname{Nazarbayev University}, \orgaddress{\street{Kabanbay Batyr Ave 53}, \city{Nur-Sultan}, \postcode{010000}, \country{Kazakhstan}}}
\affil[4]{\orgdiv{Fesenkov Astrophysical Institute}, \orgname{Nazarbayev University}, \orgaddress{\city{Almaty}, \postcode{050020}, \country{Kazakhstan}}}

%%==================================%%
%% sample for unstructured abstract %%
%%==================================%%

\abstract{The canonical approach to quantizing quantum gravity is  understood to suffer from pathological non-renomalizability.
Nevertheless in the context of effective field theory, a viable perturbative approach to calculating elementary processes is possible.
Some non-perturbative approaches, most notably loop quantum gravity and combinatorial quantum gravity imply the existence of a minimal length.
To circumvent the seeming contradiction between the existence of a minimum length and the principle of special relativity, Double Special Relativity introduces modified dispersion relationships that reconcile the conflict.
In this work, we combine these dispersion relationships with an effective field theory approach to compute the first post Newtonian correction to the bending of light by a massive object.
The calculation offers the prospect of a directly measurable effect that rests upon both the existence of a quantized gravitational field and a minimal length.
Experimental verification would provide evidence of the existence of a quantum theory of gravity, and the fundamental quantization of spacetime with a bound on the minimal distance.}

\maketitle

\section{Introduction}
Perhaps the most celebrated experimental test of General Relativity (GR) is the bending of a ray light in the presence of a gravitational field. 
For a ray of light grazing a mass $M$ at a distance $b$ from the center of mass (the impact parameter), to first order the ray is deflected by $\theta_{class}$ from straight, given by the formula \cite{ryder2009introduction},
\begin{equation}
    \theta_{class}^{(1)} =\frac{4GM}{c^2 b} \text{.}
\end{equation}
This result has been verified many times since it was first measured by Eddington in 1919 \cite{dyson1920ix}, and more recently by Bruns {\sl et al} \cite{bruns2018gravitational}, and was an early triumph of GR.

On the contrary, a quantum theory of gravity (QG) enjoys little experimental support, and indeed an accepted and consistent framing of such a theory remains elusive.
Despite many attempts to incorporate GR into a quantum field theory it remains plagued by issues of renormalizability when treated canonically \cite{dewitt1967quantum1, duff1973quantum}.
Other attempts to formulate a non-perturbative quantum theory such as loop quantum gravity and causal dynamical triangulation, amongst others, have failed to gain consensus as viable theories.
Indeed it is still a matter of some dispute as to whether a quantum theory of gravity even exists, beyond of course the well understood semi-classical treatments \cite{birrell1984quantum}.
Experimental proof of the existence of the force mediating boson of QG, the graviton, or evidence of the discreteness of spacetime geometry such as a fundamental length scale, would settle this question.

In this work we seek to bring together several independent approaches to provide a calculation of the QG correction to $\theta_{class}$, and provide a potentially measurable result that rests upon the existence of both the graviton as the quantum of the gravitational field {\sl and} a fundamental length scale consistent with a quantized spacetime.
To do so we will make use of the techniques of Effective Field Theory (EFT) \cite{donoghue1994general, bjerrum2015bending}, to compute the quantum correction to $\theta_{class}$, with the crucial added ingredient of Doubly Special Relativity (DSR) \cite{amelino2001testable,amelino2002relativity} that modifies the propagators of the graviton to account for the minimal length.
The modifications to the propagator include a coupling constant dependent upon the Planck mass, which has the effect of modifying the classical limit gravitational potential.
Our results demonstrate that the effect is just beyond the accuracy of the most recent experiments to measure light ray deflection by the Sun \cite{bruns2018gravitational}, but in principle is a measurable deflection in more distant systems such as the black hole hypothesized to exist at the center of the galaxy NGC 4395.
Further, the error bars in the measurement of the solar deflection allow us to place a lower bound on the Planck mass in the modified propagator, consistent with the theoretically proposed value of $2.176 \times 10^{-8}kg$.

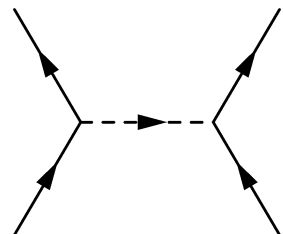
\begin{figure}[hbtp]
\centering
\begin{fmffile}{diagram}
\begin{fmfgraph*}(125,85)
    \fmfleft{i1,i2}
    \fmfright{o1,o2}
    \fmflabel{$-i g$}{v1}
    \fmflabel{$-i g$}{v2}
    \fmf{fermion,label=$p_1$}{i1,v1}
    \fmf{fermion,label=$p_1^{'}$}{v1,i2}
    \fmf{fermion,label=$p_2$}{o1,v2}
    \fmf{fermion,label=$p_2^{'}$,label.side=left}{v2,o2}
    \fmf{scalar,label=$k$}{v1,v2}
\end{fmfgraph*}
\end{fmffile}
\caption{Simplest tree level diagram in the $t$-channel ( $k=(p_1^{'}-p_1)$), for the scattering of two distinguishable fermions by the exchange of a scalar particle $\phi$ of mass $M_{\phi}$.}
\label{fig:scalarscat}
\end{figure}

Our starting point is an elementary calculation in Quantum Field Theory (QFT) that uses the Born approximation to compute the equivalent effective classical potential when comparing to a tree level calculation of the scattering amplitudes.
Following any introductory text on QFT such as \cite{peskin1995introduction}, in \Cref{fig:scalarscat} we depict the lowest order tree diagram for a process characterised by an interaction Lagrangian such as $\mathcal{L}=-g\bar{\psi} \phi \psi$.
Comparing the obtained scattering amplitude with the Born formula at first order $\sbra{ p^{'} }  iT \ket{p} = -i \tilde{V}(q)(2\pi)\delta(E_{p^{'}} - E_p)$, one obtains the expression for the potential in momentum space as,

\begin{equation*}
    \tilde{V}(\va{q}) = \frac{-g^2}{ \abs{\va{q}^2} + m^2_{\phi}} \text{.}
\end{equation*}

This may be easily inverted into spherical polar coordinate position space with the aid of a contour integral to obtain the effective ``Yukawa" potential,

\begin{equation}
    V(r)=\frac{-g^2 e^{-m_{\phi}r}}{4 \pi r} \text{.}
\end{equation}

This approach is in principle applicable to any reasonable or effective quantum field theory, such as QG providing that the perturbative expansion remains a valid approximation, as is the case in EFT.
In the case of a massless scalar with $m_\phi = 0$, one recovers the familiar inverse $r$ dependency that one finds in Newtonian gravity and electrostatics, and indeed the extension of the Feynman diagram in \Cref{fig:scalarscat} to the exchange of a massless vector gauge boson (i.e. a photon or graviton), is trivial.
Indeed in \cite{duff1973quantum,duff1974quantum,donoghue1994general}, precisely this approach was taken to obtain the classical limit of the quantum gravity potential.

\vspace{+1.5cm}

Our next ingredient is to introduce the discreteness of spacetime.
It is postulated that as a necessary consequence of the quantization of gravity, spacetime itself must be quantized and therefore discrete \cite{hossenfelder2013minimal}.
This discreteness has been extensively studied and modelled in recent studies of emergent geometry and the ground up approach to QG, combinatorial quantum gravity (CQG) \cite{trugenberger2016random, trugenberger2015quantum, trugenberger2017combinatorial,tee2020dynamics,tee2021quantum}.
These models propose that the emergence of spacetime must necessarily involve the existence of a fundamental length.
This idea ultimately has its origins in the work of Hartland Snyder \cite{snyder1947quantized}, which introduces a spatial quantum operator with a discrete spectrum (at least assuming spacetime is finite).

The existence of a fundamental length, usually assumed to the Planck length $l_p=1.6 \times 10^{-35}m$, is seemingly in direct contravention of the postulates of special relativity.
Observers in different inertial frames that are not at rest relative to each other would surely disagree about the length of the smallest quanta of distance.
However, the framework of Doubly Special Relativity (DSR) introduced by Amelino-Camelia \cite{amelino2001testable,amelino2002relativity}, provides a way out of this seemingly pathological contradiction.
In DSR \cite{jafari2020dispersion} the postulates of relativity are modified to accommodate an independent velocity and length scale, being the speed of light $c$ and the Planck length $l_p$.

Specifically for a particle of mass $m$, one modifies the normal dispersion relationship $E^2-c^2p^2 - c^4m^2=0$, with the addition of a scale dependent term $f(E,p,m;l_p)$ to give,
\begin{equation}\label{eqn:dsr_disp}
    E^2-c^2p^2 - c^4m^2 - f(E,p,m;l_p)=0 \text{.}
\end{equation}
It is possible to demonstrate that this form of modified dispersion relationship is valid in all inertial frames with an observer independent value for $l_p$, in a way that is consistent with the normal postulates of special relativity.
The leading order term, argued by dimensional analysis, has the form $f(E,p,m;l_p) \simeq l_pcEp^2$. 

A modification to the dispersion relationship implies a corresponding change to the propagator of the particle in QFT.
To be more specific, for a scalar field,  Myers {\sl et al} \cite{myers2003ultraviolet} proposed that with a preferred frame defined by a a four-vector $n^\alpha$, the normal Klein-Gordon equation for a particle of mass $m$ is replaced by,
\begin{equation}\label{eqn:kg_third_order}
    (\square + m^2 ) \Phi = \frac{ic\kappa_1}{M_p}(n \cdot \partial)^3 \Phi  \text{,}
\end{equation}
with $M_p$ being the Planck mass and $\kappa_1 $ a coupling constant.
On dimensional grounds, for $\kappa_1$ to remain dimensionless an additional factor of $c$ is necessary.
In what follows, we set $\hbar=c=1$ until we perform the explicit calculations in \Cref{sec:bending}.

This form of Equation of Motion (EOM), in momentum space would introduce a third power of the momentum, consistent with \Cref{eqn:dsr_disp}, and in fact we can generalize this approach to propose a modified dispersion relationship as follows,
\begin{equation}\label{eqn:massive_disp}
    E^2 = p^2 + m^2 + \frac{\kappa_1}{M_p} p^3 + \frac{\kappa_2}{M^2_p} p^4 \text{.}
\end{equation}
In this relationship we assume that $\kappa_1$ and $\kappa_2$ are dimensionless constants.
For a massless particle, in momentum space we of course obtain,
\begin{equation}\label{eqn:massless_disp}
    E^2 = p^2 + \frac{\kappa_1}{M_p} p^3  +\frac{\kappa_2}{M^2_p} p^4 \text{,} 
\end{equation}
to leading order.
It is possible to consider higher terms in the momentum $p$, but in this work as the propagator modifications are inversely proportional to momentum transfer we will restrict ourselves to third and fourth powers of $p$.

Defining $\alpha=\frac{1}{M_p}$, for a massless particle at constant momentum, 
one could consider the terms $\kappa_1 \alpha p^3$ and $\kappa_2 \alpha^2 p^4 $ to behave like a pseudo mass term in the EOM.
Using $m^2_e ( p )=\kappa_1 \alpha p^3 + \kappa_2 \alpha^2 p^4$  to denote this pseudo-mass term, one can rewrite the equation of motion equivalent to this dispersion relationship as,
\begin{equation}
     \Big( \square  + m^2_e (p) \Big)\Phi = 0 \text{.}
\end{equation}
The presence of this pseudo mass will affect the propagation of this particle, particularly at high momenta, but crucially will not contribute non-physical polarization states in the case of gauge bosons such as photons or gravitons. 
The particles remain massless, but their interactions will be modified by the additional terms as we shall demonstrate in \Cref{sec:mod-prop}.

This, in essence, is our approach to obtaining corrections to the Newtonian inverse square law.
Instead of a massless scalar propagator, however, we will need to consider modifications to the graviton propagator, and then use this to compute tree level scattering terms that can be compared to the Born approximation in order to extract the effective potential.

Once we have obtained our modified potential, it is straightforward to use this to compute the deviation of a ray of light that passes through this potential, and thereby compute any measurably difference to $\theta_{class}$.
We provide the details of this calculation in \Cref{sec:bending}, but our computation yields a non-zero result, which although small, could in principle be used to detect a modification to the bending of light rays from very distant objects.
As mentioned earlier we can use the error tolerance of the latest measurements of gravitational light deflection in the solar system to estimate an upper bound for $\alpha$, which we obtain numerically in the same section.
Our calculation is consistent with the definition of $\alpha$ as the inverse Planck mass.

%TO-DOs
%\begin{itemize}
  %  \item \phil{In \Cref{sec:mod-prop} introduce for massive and massless gravitons propagators for both the cubic and quartic momentum DSR terms}
  %  \item \phil{We use each of these to compute the effective potentials. This should give a matrix of 4 by 4 potentials. Key thing here is to reintroduce the factors of $\hbar$ and $c$, so we have real values for any parameters. My hope is that they look like \Cref{eqn:gen_pot} which means one light deviation calculation}
   % \item \phil{For the 4 by 4 matrix of potentials that should give us a range of values for $\alpha$ and $\beta$ in \Cref{eqn:gen_pot} and any predictions for $\theta-\theta_{class}$.}
  %  \item \phil{We could then use whatever bounds from the experimental results Nosrat found to comment on the reasonableness/scale of the result.}
%\end{itemize}

\section{The modified graviton propagator and effective potential}
\label{sec:mod-prop}
The computation of scattering amplitudes in quantum gravity was originally thought to be intractable, due to the non-renormalizability of QG.
However, it is possible within the context of an EFT \cite{donoghue1994general}, to compute a set of Feynman rules that can be applied in the normal way to scattering problems.
The benefit of EFT is to seperate out the low energy phenomena which can be consistently regularized from the high energy ones which cannot.
The approach commences by linearizing GR, by expanding the metric $g_{\mu \nu} = \eta_{\mu \nu}+ \kappa h_{\mu \nu}$ around the Minkowski background $\eta_{\mu \nu} = \text{diag}(+1,-1,-1,-1)$, with $\kappa^2=32 \pi G$.
In this scheme \cite{donoghue1994general,bjerrum2003quantum}, the relevant Feynman rules are,
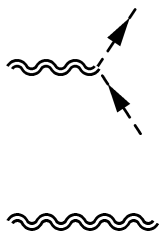
\begin{figure}[hbtp]
		\centering
        \begin{fmffile}{vrules}
        \begin{fmfgraph*}(55,50)
            \fmfright{i1,i2}
            \fmfleft{o1}
            \fmf{scalar, label=$p_1$}{i1,v1}
            \fmf{scalar, label=$p_2$}{v1,i2}
            \fmf{dbl_wiggly, label=$\alpha \beta$, label.side=top}{v1,o1}
            \fmflabel{$~~~~~=\tau^{\alpha \beta}(p_1 , p_2 , m)$, }{v1}
        \end{fmfgraph*}
        \end{fmffile}
    %\caption{Feynam Rules.}
    %\label{fig:gravscat}
\end{figure}
\vspace{-1.5cm}
\begin{figure}[hbtp]
		\centering
        \begin{fmffile}{prules}
        \begin{fmfgraph*}(55,80)
            \fmfleft{o1}
            \fmfright{o2}
            \fmf{dbl_wiggly, label=$k$}{o1,o2}
            \fmflabel{$\alpha \beta$}{o1}
            \fmflabel{$\mu \nu ~~~~~=D_{\alpha \beta \mu \nu}(k)$, }{o2}
        \end{fmfgraph*}
        \end{fmffile}
    %\caption{Feynam Rules.}
    %\label{fig:gravscat}
\end{figure}
\vspace{-1.0cm}

with the following definitions for $\tau^{\alpha \beta}$ and $D_{\alpha \beta \mu \nu}$,
\begin{align}
    \tau^{\alpha \beta}(p_1 , p_2 , m) &= -\frac{i \kappa}{2}\left \{ p_1^\alpha p_2 ^\beta + p_1^\beta p_2^\alpha - \eta^{\alpha \beta}[(p_1 \vdot p_2) - m^2]  \right\} \text{,} \\
    D_{\alpha \beta \mu \nu}(k) &= \frac{i P_{\alpha \beta, \mu \nu }}{k^2+i \epsilon}  \text{,} \label{eqn:grabprop_unmod}\\
    P_{\alpha \beta \mu \nu} &= \frac{1}{2} [ \eta_{\alpha \mu}\eta_{\beta \nu} + \eta_{\beta \mu}\eta_{\alpha \nu} - \eta_{\alpha \beta}\eta_{\mu \nu} ] \text{.}
\end{align}

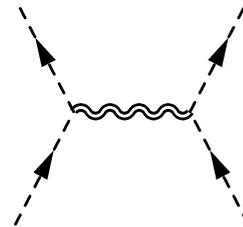
\begin{figure}[hbtp]
\centering
\subfloat{
        \begin{fmffile}{graviton}
        \begin{fmfgraph*}(110,85)
            \fmfleft{i1,i2}
            \fmfright{o1,o2}
            \fmflabel{$\tau^{\alpha \beta}$}{v1}
            \fmflabel{$\tau^{\gamma \delta}$}{v2}
            \fmf{dbl_wiggly,label=$k$}{v1,v2}
            \fmf{scalar,label=$p_1$}{i1,v1}
            \fmf{scalar,label=$p_2$}{v1,i2}
            \fmf{scalar,label=$p_3$}{o1,v2}
            \fmf{scalar,label=$p_4$,label.side=left}{v2,o2}
        \end{fmfgraph*}
        \end{fmffile}
    }
    \caption{The scattering of two massive scalars by the exchange of a graviton in the $t$-channel. The particles have mass $m_1$ and $m_2$, and the off-shell momentum $k=p_1-p_2=k_4-k_3$.}
    \label{fig:gravscat}
    \end{figure}

Using these Feynman rules we can construct a matrix element for the process depicted in \Cref{fig:gravscat}.
The contraction of the indices that results is vastly simplified by considering a non-relativistic limit where $\va{p}^2 \ll m^2$, that is the particles are slow moving, and $\va{p}$ is the three-momentum of the particle.
Following \cite{bjerrum2015bending,bjerrum2016light} we obtain for the non-relativistic matrix element $M_{12}(p_1,p_3 \rightarrow p_2,p_4)$ of \Cref{fig:gravscat}, appropriately normalized by $\frac{1}{2m_1 2m_2}$, 

\begin{equation}\label{eqn:matrix_el}
    iM_{12}(p_1,p_3 \rightarrow p_2,p_4)= i\frac{1}{2m_1 2m_2}\mathcal{M}_{12} = -\frac{\kappa^2}{8}\frac{m_1 m_2}{\va{k}^2} \text{.}
\end{equation}
We compare this with the Born approximation for the same scattering process to extract an effective potential, after Fourier transform as follows,
\begin{align}
    V(\va{r}) &= -\frac{\kappa^2 m_1 m_2}{8} \int\limits_{0}^{\infty} \frac{d^3 \va{k}}{(2\pi)^3} \frac{e^{i \va{k} \cdot \va{r}} }{\va{k}^2} \text{,} \label{eqn:fourier_pot}\\
    &= -G \frac{m_1 m_2}{r} \text{.}
\end{align}

Our approach is to modify $D_{\alpha \beta \mu \nu}(k)$ to account for the modified dispersion relationships from DSR \cite{amelino2002relativity}, as outlined in \Cref{eqn:massive_disp,eqn:massless_disp}.
We consider the massless case.
We have essentially two choices, one which is cubic in the exchanged momentum $\va{k}$, and a further option which is quartic.
With $\alpha=\frac{1}{M_p}$, we set $\kappa_1=\kappa_2=1$ in  \Cref{eqn:massless_disp}, noting that we can reinsert a non unity value for the constants in our calculations later. 
This leads to two alternatives for the propagator.
\begin{align}
    iD_{\mu \nu \alpha \beta}(k) &= \frac{i}{k^2( 1 + i\alpha k)} P_{\mu \nu  \alpha \beta} \label{eqn:mod_cubic} \text{,} \\
    iD_{\mu \nu \alpha \beta}(k) &= \frac{i}{k^2( 1 + \alpha^2 k^2)} P_{\mu \nu  \alpha \beta} \label{eqn:mod_quartic} \text{.}
\end{align}
The factor of $i$ in \Cref{eqn:mod_cubic} suggested by the imaginary right hand side of \Cref{eqn:kg_third_order}, is in fact necessary for this propagator to yield a physically real potential.

We substitute both of these propagators into \Cref{eqn:matrix_el} to extract an expression for the classical potential by comparison with the Born approximation.
The details of the computation of the Fourier transform are in the appendix, but they both rely upon converting the integral in \Cref{eqn:fourier_pot} into spherical polar $\va{k}$ space.
Once this is performed, we are left with an integral over $k=\abs{\va{k}}$.
In the case of the quartic propagator \Cref{eqn:mod_quartic}, we obtain,
\begin{equation}\label{eqn:quartic_pot}
    V(r) = -G \frac{m_1 m_2}{r} \left( 1 - e^{- \frac{r}{\alpha}} \right ) \text{,}
\end{equation}
the details of the computation being described in \Cref{sec:appendix_a}.
For the cubic denominator in \Cref{eqn:mod_cubic}, one obtains a very similar result for the modified potential (details in \Cref{sec:appendix_b}),
\begin{equation}\label{eqn:cubic_pot}
    V(r) = -G \frac{m_1 m_2}{r} \left( 1 - 2 e^{- \frac{r}{\alpha}} \right ) \text{.}
\end{equation}
These potentials have a common general form,
\begin{equation}\label{eqn:gen_pot}
    V(r) = -G \frac{m_1 m_2}{r} \left( 1 + \beta e^{- \lambda r} \right ) \text{,}
\end{equation}
with $\beta$ and $\lambda$ representing both the strength ($\beta$) and range $\lambda^{-1}$ of the correction to the Newtonian potential, a convention followed in \cite{henrichs2021testing}.

\section{Bending of gravitational waves around a massive object}
\label{sec:bending}
One of the earliest tests of GR was the bending of light by an angle $\theta_{class}$ around a massive object, with the classical result for first order post-Newtonian approximation given by,
\begin{equation}\label{eqn:class_bend}
    \theta_{class}^{(1)} =\frac{4GM}{c^2 b} \text{,}
\end{equation}
where $b$, the impact parameter, is the closest distance to the center of the mass $M$ of the light ray \cite{ryder2009introduction}. 
This result is obtained as a perturbative expansion of a geodesic computation in the Schwarzchild metric \cite{bjerrum2016light}, and this is the first term.
The second term in the expansion of $\order{G^2}$ is given by,
\begin{equation}\label{eqn:class_dev_second}
    \theta_{class}^{(2)} = \frac{15\pi G^2 M^2}{4 c^4 b^2} \text{.}
\end{equation}
Our computation in this section is firmly first order in the gravitational coupling, but we will also compare our results to \Cref{eqn:class_dev_second} as well as the first order correction.

The bending of light has been analyzed using quantum gravity in the effective field theory treatment \cite{bjerrum2003quantum,bjerrum2015bending,bjerrum2016light} where it is has been shown that the first two post Newtonian terms are obtained.
The first QG correction is obtained at one-loop, but as we shall see our modified propagator actually affects the tree level contribution.
The deflection angle for a ray of light is computable for a general potential $V(r)$, by evaluating the following integral,
\begin{equation}\label{eqn:angular_div}
    \theta= -\frac{b}{\hbar \omega} \int\limits_{-\infty}^{\infty} V^{'}(b\sqrt{1+u^2})\frac{\dd u}{\sqrt{1+u^2}} \text{.}
\end{equation}
where $\omega$ is the frequency of the radiation, and $u=\frac{ct}{b}$ is a convenient integration variable, introduced to simplify the computation.

Our strategy is to use the potential in \Cref{eqn:gen_pot} to solve the above equation and determine the first order correction to the classical result.
Although this potential is obtained by the scattering of massive scalar particles, we obtain a general form for the classical limit to the potential.
In \cite{bjerrum2016light}, a much more detailed computation using classical propagators is undertaken, but our computation is at the tree level only. 
Accordingly we compare our result to the first order post-Newtonian result in \Cref{eqn:class_bend}.
Substituting \Cref{eqn:gen_pot} into \Cref{eqn:angular_div} and performing the relevant integration, we obtain our result in terms of the Bickley-Naylor functions $Ki_n(x)$ \cite{altacc2007exact},
\begin{equation}\label{eqn:final_dev}
    \theta-\theta_{class} = \frac{4GM\beta}{c^2 b} Ki_2 \left (\frac{b}{\alpha} \right ) + \frac{4GM\beta}{c^2 \alpha} Ki_1 \left (\frac{b}{\alpha} \right ) \text{.}
\end{equation}
The details of this calculation are in \Cref{sec:angular_calc}.

The Bickley-Naylor functions have an asymptotic form \cite{milgram1978analytic} form which can allow us to compute an order of magnitude calculation for \Cref{eqn:final_dev}.
In particular for $x \gg 1$, we have,
\begin{equation}
    Ki_n(x) \approxeq \sqrt{ \frac{\pi}{2x} } e^{-x} \left [1 - \frac{(1+4n)}{8x} + \order{\frac{1}{x^2}} \dots \right] \text{.}
\end{equation}
In our calculation $b \sim 7 \times 10^8 \text{m and } \alpha \sim 10^{-8}\text{kg}$, and so we can neglect all of the terms in the square brackets.
This simplifies \Cref{eqn:final_dev}, as for $x \gg 1$, we note that $Ki_1(x) = Ki_2(x)$.
Collecting terms we arrive at the asymptotic estimate for the correction as,
\begin{equation}\label{eqn:approx_delta}
   \theta-\theta_{class} = \frac{1}{c^2}\sqrt{ \frac{\pi \alpha}{2b} } 4GM\beta e^{-b/\alpha} \left( \frac{b + \alpha}{b \alpha}\right ) \text{.}
\end{equation}
If we imagine a light ray grazing the sun we can compute the order of magnitude of this deflection, assuming $\alpha=1/M_p = 4.587 \times 10^7 kg$, and $b=6.9 \times 10^8 m$. 
On dimensional grounds we need to reinsert the values for $\kappa_1$ and $\kappa_2$ from \Cref{eqn:massive_disp,eqn:massless_disp} to convert the $\alpha$ to a length, and obtain an angle of deflection.
The presence of the exponential term in this expression, however, would give a vanishingly small correction for any value of $\kappa_{1}$ or $\kappa_2$ that is much smaller than $1$, and indeed the approximation requires that the ratio $b/\alpha \gg 1$, so for the purposes of estimation we retain their value at unity.
For completeness, however, we can quickly see that if $\Delta \theta = \theta-\theta_{class}$, we can extract the dependency of the result on $\kappa$ in the following manner,
\begin{align}
    \Delta_{\kappa \neq 1}( \theta ) & = f(\kappa,b,\alpha) \Delta_{\kappa=1}( \theta ) \text{,}\\
    f(\kappa,b,\alpha) &= \frac{b+\kappa \alpha}{\sqrt{\kappa}(b+\alpha)} e^{\frac{b(\kappa-1)}{\kappa \alpha } } \text{.}
\end{align}
Insertion of the estimates above, we obtain for the deflection, $\Delta_{\kappa=1}(\theta) = -1.2998\times10^{-11} \beta$ radians, with $\beta=-1$ for the quartic propagator and $\beta=-2$ for the cubic.
It should also be remarked that for values of $\kappa < 1$, the scaling function $f(\kappa,b,\alpha)$ decays to zero extremely rapidly, due in large part to the exponential, and also that our asymptotic expansion of the Bickley-Naylor functions become unreliable. 
It would seem that prospects for detection of such a deviation are remote, given that in the most recent measurements of light deflection by Bruns {\sl et al} \cite{bruns2018gravitational} during the 2017 solar eclipse had an overall error of $1.536 \times 10^{-8}$ radians.
However, the prospects of obtaining a meaningful deviation are greatly improved when considering much more massive bodies than the Sun.
The term which dominates the correction to the angular deflection is the exponential $e^{-b/\alpha}$, which for $b/\alpha \gg 1$ becomes infinitesimally small.
As previously noted, the integrity of our approximation requires it remains significantly greater than one, but it is interesting to consider what physical systems would give a ratio that approaches unity.
If instead of a star, we consider a black hole, we can set the parameter $b$ to be the Schwarzchild radius. 
The ratio of $b/\alpha$ is equal to one for a black hole of mass $M=1.553 \times 10^4  M_\odot$, indicating that we should look at the category of  intermediate mass black holes.
Such a candidate would be the black hole believed to inhabit the galaxy NGC 4395, which has a consensus mass value of $3.6 \times 10^5 M_\odot$~ \cite{peterson2005multiwavelength}, although other estimates place it somewhat lower.
The ratio has a value of $b/\alpha=23.185$ if we set $\kappa=1$, and is at least one order of magnitude larger than $1.0$, and so our approximation should have reasonable validity.
Using the formula \Cref{eqn:approx_delta}, we obtain a value of $\Delta_{\kappa=1}(\theta) = -1.942 \times 10^{-9}$ radians, for a value of $b/\alpha=23.185$ if we set $\kappa=1$.
If we push the limits of the approximation of the Bickley-Naylor functions and consider a black hole of $M=-1.553 \times 10^4  M_\odot$, we must abandon the approximation and instead use a table of computed values for the Bickley-Naylor functions (see for example \cite{abramowitz1964handbook} pg 489 for details on how to compute exact values, or \cite{bickley1935xxv} for tabular values).
When we do this the deviation is $-0.383 \pi$ radians, or $-68.97^{\circ}$ degrees!
This does indicate that there may well be intermediate sized black holes for which the quantum correction to the leading order classical light deviation becomes appreciable, and in principal measurable.

Alternatively we can consider \Cref{eqn:final_dev} for the solar system, and ask what is the maximum value of $\alpha$ that would produce a deflection that is {\sl not} measurable given current experimental constraints.
This amounts to solving the equation numerically for the value of $\alpha$ such that $\Delta \theta \leq 1.536 \times 10^{-8}$ radians from the Bruns {\sl et al} study \cite{bruns2018gravitational}.
This numerical solution can be obtained by a combination of Newton's method and a fine tuned iteration of $alpha$ using standard methods (a short program implemented in python is available from the authors on request).
Intriguingly the result is equivalent to a value of $M_g \geq 1.093 \times 10^{-8} kg$, which is within the accuracy of the method one-half the theoretical value of the Planck mass $2.176 \times 10^{-8}kg$.

We summarize these results in \Cref{tab:deflect_res}, and include the second order post-Newtonian corrections from \Cref{eqn:class_dev_second}.
We should treat the computations for light bending around the black holes considered somewhat carefully as the perturbative computation requires $\frac{2GM}{c^2 b} \ll 1$, and for both of our bodies we have $0.995$ for NGC 4395 and $1.000$ for our extremal black hole. 
Nevertheless, we note that in all cases our quantum correction is smaller than the classical one, with the extremal black hole yielding a result of similar size.
Although the correction we obtained for the bending of light around the Sun is well below current experimental bounds, perhaps in the era on non-earth bound astronomy it is conceivable that this effect could be measured.

\begin{table*}[hbtp]
\centering
\begin{tabular}{|c|c|c|c|c|}
\hline
\rule{0pt}{10pt} Gravitating Body & GR $\order{G}$ & GR $\order{G^2}$ & DSR Calculation \rule{0pt}{10pt}  \\ \hline
\rule{0pt}{10pt} Sun ($b=6.951\times10^8m$) & $8.493\times10^{-6}$ & $5.311 \times 10^{-11}$ & $-1.299\times10^{-11}$ \rule{0pt}{10pt} \\  \hline 
\rule{0pt}{10pt} NGC 4395 ($b=1.069\times10^9m$) & $1.989$ & $2.915$ & $-1.045\times10^{-9}$ \rule{0pt}{10pt} \\ \hline
\rule{0pt}{10pt} Extremal Black Hole ($b=4.587\times10^7m$) & $1.999$ & $2.945$ & $-1.204$  \rule{0pt}{10pt} \\ \hline
\end{tabular}
\caption{The deflection of light across a selection of gravitating bodies. The impact parameter in the case of the Sun is its radius, and for the two Black Holes their respective Schwarzchild radii. All results are stated in radians and the values are computed using equations \Cref{eqn:class_bend}, \Cref{eqn:class_dev_second} and \Cref{eqn:final_dev}.}
\label{tab:deflect_res}
\end{table*}

\section{Conclusion and Discussions}
\label{sec:conclusion}

In this work we have applied the modified propagators of DSR to the EFT treatment of quantum gravity.
The modification of the propagator has an impact on the very high energy short distance behavior of gravity, but however does modify the EFT computation of the classical limit gravitational potential.
Using this modification we have successfully computed the correction to the first post-Newtonian term in the deflection of a light ray, well understood from General Relativity.
These corrections are stated in terms of the Bickley-Naylor functions in \Cref{eqn:final_dev}.

Unsurprisingly when applied to the Sun, these corrections are well beyond the sensitivity of even the most recent measurements.
However, when we consider much more massive and dense objects, such as intermediate mass black holes, the corrections are much larger. 
Indeed for a Black Hole of mass $1.5\times10^4 M_\odot$ we obtain a correction of $-68.97^{\circ}$ degrees.
For the solar system the error in the Bruns {\sl et al} study allows us to place a bound on the coupling constants introduced in our modified propagator.
By numerical calculation we can show that a coupling in the order of $\alpha=\frac{1}{M_p}$ is consistent with the error tolerance of the experimental result.

Our calculation, of course, has many limitations. 
Not least of which is the propagator we derived from DSR when used with the other components of the Feynman diagram machinery will cause the vertices to cease to observe Lorentz invariance. 
Building a comprehensive and consistent set of EFT Feynman rules incorporating the modified propagator is the subject of further work, but is an acknowledged drawback.
Further, in terrestrial applications, the result for the variance form the first order correction is not significant enough to be measurable.
We do however believe that the result for black holes, particularly in the era of non-terrestrial astronomy does at least hold out the prospect of being measurable.

Given that the result rests both on the assumption of a quantized gravity field, and a minimum length, validation of the existence of a quantum theory of gravity would be achieved should the calculations in the work achieve experimental confirmation.
We accept that the result is somewhat beyond feasibility at this stage, but at the very least it is not in conflict with current experimental evidence.

%\begin{equation}
%    -\frac{8\alpha^5k^2}{(1+\alpha^2k^2)^2} + \frac{2\alpha^3}{(1+\alpha^2k^2)^2}
%\end{equation}
\appendix
\section{Computation of Effective Potential - Quartic Propagator}
\label{sec:appendix_a}

Our starting point to compute \Cref{eqn:mod_four_pot}, is \Cref{eqn:fourier_pot} suitably modified by inserting the modified propagator \Cref{eqn:mod_quartic}.
Performing the substitution we have,
\begin{equation}\label{eqn:mod_four_pot}
    V(\va{r}) = - \frac{\kappa^2 m_1 m_2}{8} \int\limits_{0}^{\infty} \frac{d^3 \va{k}}{(2\pi)^3} \frac{e^{i \va{k} \cdot \va{r}} }{\va{k}^2 ( 1 + \alpha^2 \va{k}^2)} \text{.}
\end{equation}

We can approach this integral by converting to spherical polar coordinates in momentum space $(k,\theta,\varphi)$, obtaining,
\begin{align*}
    V(\va{r}) &= -\frac{\kappa^2 m_1 m_2}{8} I(k) \text{,} \\
    I(k) &= \frac{1}{(2\pi)^3} \int\limits_{0}^{2\pi} \int\limits_{0}^{\pi} \int\limits_{0}^{\infty} \frac{e^{irk \cos{\theta}}}{k^2(1+\alpha^2 k^2)} k^2 \sin^2{\theta} ~\dd k ~\dd \theta ~\dd \varphi  \text{.}
\end{align*}
The angular integrals can be performed, after making the substitution $z=\cos{\theta}$, obtaining the following integral in $k$,
\begin{equation}
    I(k)=\frac{1}{(2\pi r)(2 \pi i)} 2 \int\limits_{0}^{\infty} \frac{k e^{ikr}}{k^2(1+\alpha^2 k^2)} ~\dd k \text{.}
\end{equation}
As the integrand is even, we can extend the limits $2\int\limits_{0}^{\infty} = \int\limits_{-\infty}^{\infty}$, leaving the following integral to be performed,

\begin{equation}\label{eqn:final_int}
    I(k)=\frac{1}{(2\pi r)(2 \pi i)} \int\limits_{-\infty}^{\infty} \frac{k e^{ikr}}{k^2(1+\alpha^2 k^2)} ~\dd k \text{.}
\end{equation}

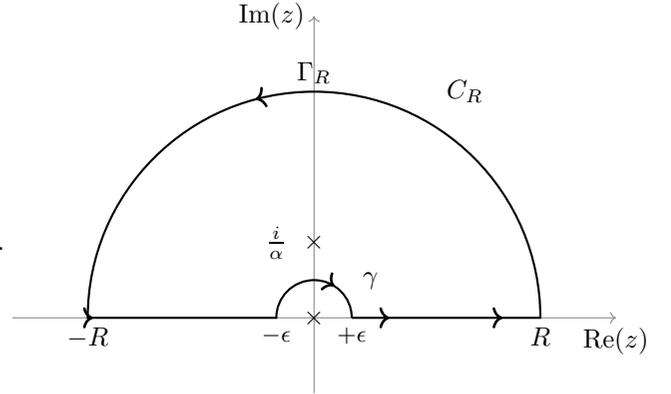
\begin{figure}[hbtp]
    \centering
    \begin{tikzpicture}
        [
        decoration={%
        markings,
        mark=at position 0.5cm with {\arrow[line width=1pt]{>}},
        mark=at position 2cm with {\arrow[line width=1pt]{>}},
        mark=at position 0.5 with {\arrow[line width=1pt]{>}},
        mark=at position 0.75 with {\arrow[line width=1pt]{>}},
        mark=at position -5mm with {\arrow[line width=1pt]{>}},
        }
    ]
    \draw [help lines,->] (-4,0) -- (4,0) coordinate (xaxis);
    \draw [help lines,->] (0,-1) -- (0,4) coordinate (yaxis);
    \node at (0,1) {$\times$};
    \node at (-.5,1) {$\frac{i}{\alpha}$};
    \node at (0,0) {$\times$};
    \path [draw, line width=0.8pt, postaction=decorate] (0.5,0) node [below, font=\scriptsize] {$+\epsilon$} -- (3,0) node [below] {$R$} arc (0:180:3) node [below] {$-R$} -- (-0.5,0) node [below, font=\scriptsize] {$-\epsilon$} arc (180:0:.5);
    \node [below] at (xaxis) {$\operatorname{Re}(z)$};
    \node [left] at (yaxis) {$\operatorname{Im}(z)$};
    %\node [above left] {$O$};
    \node at (0.75,0.5) {$\gamma$};
    %\node at (1,.8) {$C_{\varepsilon}$};
    \node at (2,3) {$C_{R}$};
    \node at (0,3.25) {$\Gamma_{R}$};

    \end{tikzpicture}
    \caption{Contour used to integrate the Fourier transformed potential in \Cref{eqn:final_int}. We note that the complex function has poles at $0$ and $\pm \frac{i}{\alpha}$.}
    \label{fig:quartic_contour}
\end{figure}

To perform this integral, we consider the contour in the complex plane in \Cref{fig:quartic_contour}, of the complex function,
\begin{equation}\label{eqn:complex_fun}
    f(z)=\frac{z e^{irz}}{\alpha^2 z^2(z^2 + \frac{1}{\alpha^2})} \text{,}
\end{equation}

This function \Cref{eqn:complex_fun} has poles at $0$ and $\pm\frac{i}{\alpha}$, with the pole at $+\frac{i}{\alpha}$ being included in the contour $C_R$.
These poles have residues of $1$ and $-\frac{1}{2}e^{-r/\alpha}$ respectively.
The integral around the contour can be split as follows,
\begin{equation*}
    \int_{C_R} = \int\limits_{-R}^{-\epsilon} + \int_\gamma + \int\limits_{\epsilon}^R + \int_{\Gamma_R} = 2\pi i (-\frac{e^{-r/\alpha}}{2})\text{.}
\end{equation*}
If we allow $R\rightarrow \infty$ and $\epsilon \rightarrow 0$, the integral $\int_\gamma \rightarrow - \pi i$ (noting that the half semicircle contour $\gamma$ is traversed in the clockwise direction), and the integrals $\int\limits_{-R}^{-\epsilon} + \int\limits_{\epsilon}^R \rightarrow \int\limits_{-\infty}^{\infty}$, whilst $\int_{\Gamma_R} \rightarrow 0$.

We therefore obtain the following value for the integral,
\begin{equation*}
    \int_{-\infty}^{\infty} \frac{ke^{irk}}{k^2(1+\alpha^2k^2)} ~\dd k = \pi i( 1 - e^{-\frac{r}{\alpha}} )
\end{equation*}

If we substitute this back into \Cref{eqn:mod_four_pot}, and expand $\kappa^2$, we obtain our final result,

\begin{equation}
    V(r) = -G \frac{m_1 m_2}{r} \left( 1 - e^{- \frac{r}{\alpha}} \right ) \text{.}
\end{equation}

\section{Computation of Effective Potential - Cubic Propagator}
\label{sec:appendix_b}
The computation here is essentially identical to that for the quartic propogator, but it results in the integral of the following compex function around a contour identical to that draw in \Cref{fig:quartic_contour},
\begin{equation}\label{eqn:cubic_complex_fun}
    f(z)=\frac{z e^{irz}}{z^2(1+i \alpha z)} \text{.}
\end{equation}
This function has poles at $z=0$ and $z=\frac{i}{\alpha}$, with residues of $1$ and $-\alpha e^{-r/\alpha}$ respectively.
Following through the same analysis as in \Cref{sec:appendix_a}, one obtains for the effective potential for the cubic propagator,

\begin{equation}
    V(r) = -G \frac{m_1 m_2}{r} \left( 1 - 2 e^{- \frac{r}{\alpha}} \right ) \text{.}
\end{equation}

\section{Computation of the quantum correction from our modified propagator to the lensing of gravitational waves around a massive object}
\label{sec:angular_calc}

Our starting point is \Cref{eqn:gen_pot}, which when we differentiate with respect to $r$ we obtain the following three contributions,
\begin{equation}\label{eqn:pot_deriv}
    \dv{V(r)}{r} = -\frac{A}{r^2} -\frac{A\beta e^{-r/\alpha}}{r^2} - \frac{A\beta e^{-r/\alpha}}{\alpha r} \text{,}
\end{equation}
with $A=2GM m_g$, and $M$ is the mass of the object the light ray is grazing, and $m_g$ is the mass of the graviton.

Performing the substitution $r \rightarrow b\sqrt{1+u^2}$, and inserting into \Cref{eqn:angular_div}, we are left with three integrals to perform,
\begin{align}
    \theta &=\frac{b}{\hbar \omega}\left ( I_1 +I_2 +I_3 \right )  \text{,} \label{eqn:adiv}\\
    I_1 &= \frac{A}{b^2} \int\limits_{-\infty}^{\infty} \frac{\dd u}{(1+u^2)^{3/2}} \label{eqn:I1}\\
    I_2 &= \frac{A\beta}{b^2} \int\limits_{-\infty}^{\infty} \frac{e^{-C \sqrt{1+u^2}} }{(1+u^2)^{3/2}} ~\dd u \text{,} \label{eqn:I2} \\
    I_3 &= \frac{A\beta}{\alpha b} \int\limits_{-\infty}^{\infty} \frac{e^{-C \sqrt{1+u^2}} }{1+u^2} ~\dd u \text{,} \label{eqn:I3} \\
    C &= \frac{b}{\alpha} \text{.}
\end{align}
The first integral is elementary and may be solved by making the substitution $u=\tan \theta$, which results in an integral $\int\limits_{-\pi/2}^{\pi/2} \cos{\theta} ~\dd \theta = 2$. 

The second integral is however more challenging, and requires the use of Bickley-Naylor special functions \cite{altacc2007exact}, first studied as solutions to certain thermal radiation problems.
To make progress we note that the integrand is analytic everywhere on the real line with constant limits.
As such we are free to differentiate under the integral sign with respect to $C$, obtaining (ignoring the leading factors which we will reintroduce later),
\begin{equation}\label{eqn:dI2}
    \dv{I_2 (C)}{C} = - I_4 = - \int\limits_{-\infty}^{\infty} \frac{e^{-C \sqrt{1+u^2}}}{1+u^2} ~\dd u \text{.}
\end{equation}
It will be noted that this is identical in form to \Cref{eqn:I3}.
This can be further simplified by making the substitution $x=\sinh (u)$, which yields the following integral,
\begin{equation*}
    I_4=2\int_0^{\infty} \frac{e^{-C \cosh (x)}}{\cosh (x)} ~\dd x\text{.}
\end{equation*}
This equation is the definition of the Bickley-Naylor function $Ki_n(C)$ of order $n=1$.
There is an extremely convenient differential relation for the Bickley-Naylor functtions,
\begin{equation}
    \dv{Ki_{n+1}(x)}{x} = - Ki_n(x) \text{,}
\end{equation}
that allows us to immediately solve our differential equation for $I_2(C)$, to obtain
\begin{equation}
    I_2=2 Ki_2(C) + \text{const} \text{.}
\end{equation}
The constant of integration can be determined to be zero by noting that for $C=0$, $I_2 = 2$, and that $Ki_2(0) = 1$.

The work invested in computing our second integral, involved solving $I_3$ as an intermediate step, and so we can conclude $I_3=\frac{2 A\beta\pi}{\alpha b}Ki_1(C)$ without further work.

Bringing all of our results together and substituting for $C$, we have,
\begin{align*}
    &\theta=\frac{b}{\hbar \omega}(I_1+I_2+I_3) = \\
    & \frac{1}{\hbar \omega}\left [ \frac{2A}{b} + \frac{2A\beta}{b} Ki_2 \left (\frac{b}{\alpha} \right ) + \frac{2A\beta}{\alpha} Ki_1 \left (\frac{b}{\alpha} \right )\right ]\text{.}
\end{align*}
We note that the factor $\hbar \omega$ is the energy of the gravitons, which is equivalent to the factor $m_g$, as we are working in units of $c=1$.
Substituting this result into \cref{eqn:adiv}, we obtain our final result,

\begin{align}
    \theta&=\frac{4GM}{b} + \\
    & \frac{4GM\beta}{b} Ki_2 \left (\frac{b}{\alpha} \right ) + \frac{4GM\beta}{\alpha} Ki_1 \left (\frac{b}{\alpha} \right ) \text{,}
\end{align}
and note that the leading term is the first contribution to the classical deviation term obtainable from GR, $\theta_{class}$.

%\tableofcontents

\section*{Acknowledgements}
The authors would like to thank Prof. Michael Bluck for his invaluable assistance with Appendix C, and in particular bringing the Bickley-Naylor functions to our attention. In addition we are grateful for the input from Prof. Paul Davies and Prof. Giovanni Amelino-Camelia whose many insights have proven invaluable.

PT acknowledges the support of the Royal Society grant EiR\textbackslash211364. 
NJ has been funded by the Science Committee of the Ministry of Education and Science of the Republic of Kazakhstan (Grant No. AP13869354)

%%===========================================================================================%%
%% If you are submitting to one of the Nature Portfolio journals, using the eJP submission   %%
%% system, please include the references within the manuscript file itself. You may do this  %%
%% by copying the reference list from your .bbl file, paste it into the main manuscript .tex %%
%% file, and delete the associated \verb+\bibliography+ commands.                            %%
%%===========================================================================================%%
%\clearpage
\bibliographystyle{sn-mathphys}
\bibliography{DSRAndLight-EPJC}% common bib file
%% if required, the content of .bbl file can be included here once bbl is generated
%%\input sn-article.bbl

%% Default %%
%%\input sn-sample-bib.tex%

\end{document}